\documentclass[conference]{IEEEtran}
\IEEEoverridecommandlockouts
\usepackage{cite}
\usepackage{amsmath,amssymb,amsfonts}
\usepackage{algorithmic}
\usepackage{graphicx}
\usepackage{textcomp}
\usepackage{xcolor}
\usepackage[utf8]{inputenc}

\usepackage{url}
\usepackage{subcaption}
\usepackage{graphicx}
\usepackage{scrextend}
\usepackage{tikz}
\usepackage{tkz-tab}
\usepackage{multirow}
\usepackage{listings}
\usepackage{latexsym}
\usepackage{amssymb}
\usepackage{amsmath}
\usepackage{subcaption}
\usepackage{mathtools}
\usepackage[]{moresize}
\usepackage{soul}
\usepackage{enumitem}   

\usepackage{tikz}
\usepackage{tkz-tab}
\usetikzlibrary{automata,arrows,positioning,calc}
\usetikzlibrary{shapes,snakes}

\usepackage{tabularx,booktabs}
\newcolumntype{C}{>{\centering\arraybackslash}X} 
\newcolumntype{b}{X}
\newcolumntype{s}{>{\hsize=.5\hsize}X}
\newcolumntype{v}{>{\hsize=.3\hsize}X}

\DeclarePairedDelimiter{\ceil}{\lceil}{\rceil}

 \pdfoutput=1

\def\BibTeX{{\rm B\kern-.05em{\sc i\kern-.025em b}\kern-.08em
		T\kern-.1667em\lower.7ex\hbox{E}\kern-.125emX}}
\begin{document}
	
	\title{Komondor: a Wireless Network Simulator for Next-Generation High-Density WLANs\\
		\thanks{This work has been partially supported by a Gift from CISCO University Research Program (CG\#890107) \& Silicon Valley Community Foundation, by the Spanish Ministry of Economy and Competitiveness under the Maria de Maeztu Units of Excellence Programme (MDM-2015-0502), and by the Catalan Government under grant SGR-2017-1188. The work by S. Barrachina-Mu\~noz is supported by an FI grant from the Generalitat de Catalunya.}
	}
		
	\author{\IEEEauthorblockN{1\textsuperscript{st} Sergio Barrachina-Muñoz}
		\IEEEauthorblockA{\textit{Wireless Networking (WN)} \\
			\textit{Universitat Pompeu Fabra}\\
			Barcelona, Spain \\
			sergio.barrachina@upf.edu}
		\and
		\IEEEauthorblockN{2\textsuperscript{nd} Francesc Wilhelmi}
		\IEEEauthorblockA{\textit{Wireless Networking (WN)} \\
			\textit{Universitat Pompeu Fabra}\\
			Barcelona, Spain \\
			francisco.wilhelmi@upf.edu}
		\and
		\IEEEauthorblockN{3\textsuperscript{rd} Ioannis Selinis}
		\IEEEauthorblockA{ \textit {Inst. for Com. Systems 5GIC} \\
			\textit{University of Surrey}\\
			Guildford, United Kingdom \\
			ioannis.selinis@surrey.ac.uk}
		\and
		\IEEEauthorblockN{4\textsuperscript{th} Boris Bellalta}
		\IEEEauthorblockA{\textit{Wireless Networking (WN)} \\
			\textit{Universitat Pompeu Fabra}\\
			Barcelona, Spain \\
			boris.bellalta@upf.edu}
	}
	
	\maketitle

	\begin{abstract}
		
		Komondor is a wireless network simulator for next-generation wireless local area networks (WLANs).
		The simulator has been conceived as an accessible (ready-to-use) open source tool for research on wireless networks and academia.
		An important advantage of Komondor over other well-known wireless simulators lies in its high event processing rate, which is furnished by the simplification of the core operation. This allows outperforming the execution time of other simulators like ns-3, thus supporting large-scale scenarios with a huge number of nodes.
		In this paper, we provide insights into the Komondor simulator and overview its main features, development stages and use cases. The operation of Komondor is validated in a variety of scenarios against different tools: the ns-3 simulator and two analytical tools based on Continuous Time Markov Networks (CTMNs) and the Bianchi's DCF model. Results show that Komondor captures the IEEE 802.11 operation very similarly to ns-3. Finally, we discuss the potential of Komondor for simulating complex environments -- even with machine learning support -- in next-generation WLANs by easily developing new user-defined modules of code.
	\end{abstract}
	
	\begin{IEEEkeywords}
		Wireless network simulator, high-density, WLAN, IEEE 802.11ax, machine learning 
	\end{IEEEkeywords}
	
	\section{Introduction}
	\label{section:introduction}
	
\begin{figure}[t]
		\centering	
		\includegraphics[width=0.75\columnwidth]{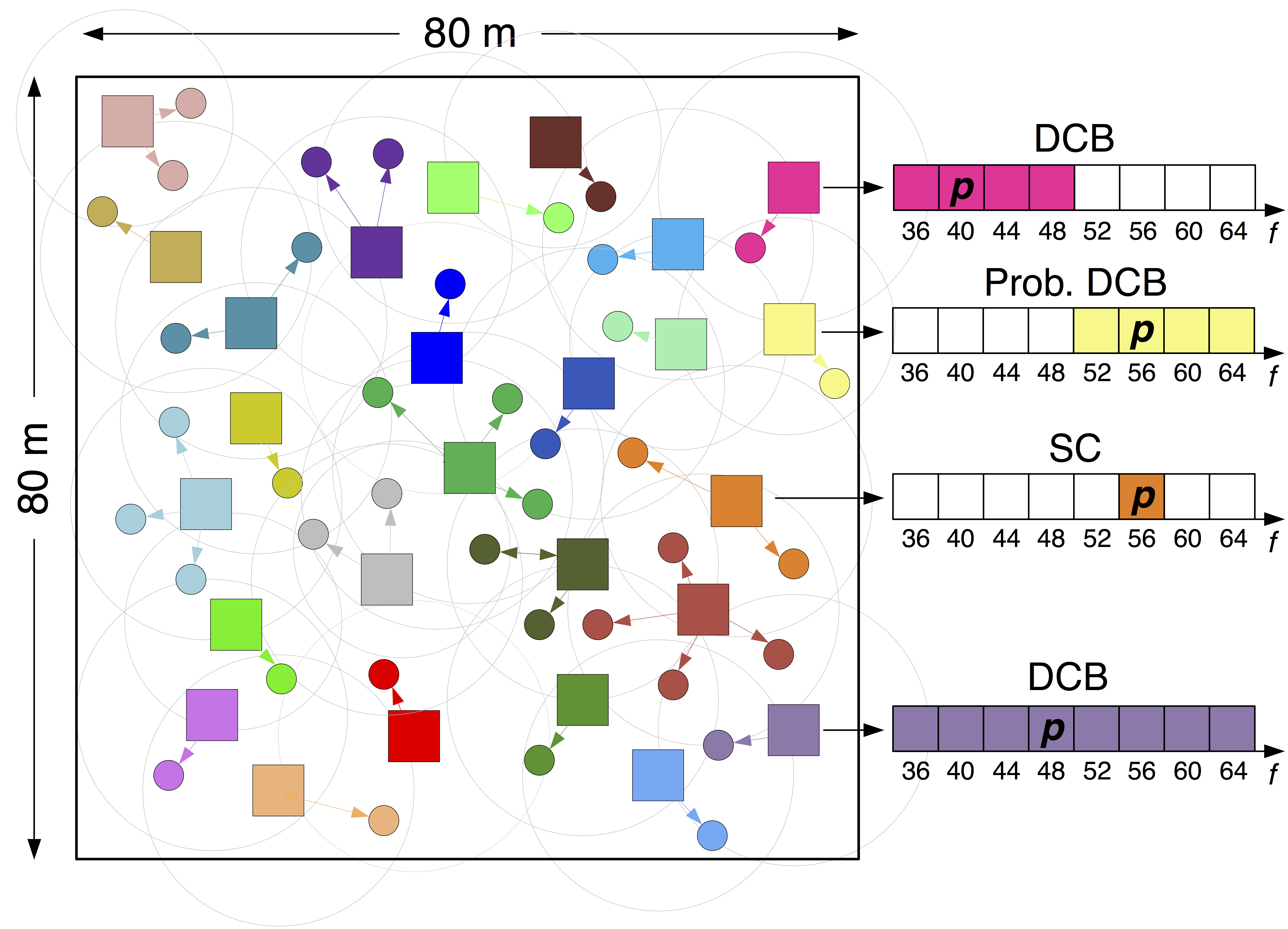}
		\caption{Dense scenario composed of 25 WLANs. Note that each WLAN has its own channel allocation and DCB policy.}
		\label{fig:map_dense}
	\end{figure}

	The Institute of Electrical and Electronics Engineers (IEEE) 802.11 Wireless Local Area Networks (WLANs) are evolving fast to satisfy the new strict requirements in terms of data rate and user density. In particular, various IEEE 802.11 amendments have been introduced in the past few years or are under active development to accommodate the need for higher capacity, exponential growth in number of devices, and novel use-cases.~\cite{8485317_selOA}. An example of next-generation high-density deployment is depicted in Fig. \ref{fig:map_dense} where multiple WLANs are allocated with different channels and dynamic channel bonding (DCB) policies.
	
	Of particular interest is the IEEE 802.11ax (11ax) amendment~\cite{tgax2017draft}, that is under active development and which was introduced to address the demands and challenges that WLANs will face in the congested 2.4/5 GHz bands~\cite{7422404_bellalta11ax}. Other important amendments for next-generation wireless networks are the IEEE 802.11ay \cite{ghasempour2017ieee} and EXtreme Throughput (XT) 802.11 \cite{xtreme_throughput}, which aim to exploit the 60 GHz and $\leq 6$ GHz frequency bands, respectively. Amendments like the aforementioned ones lay the foundation of next-generation WLANs by including new features such as multiple-antenna techniques like Downlink/Uplink Multi-User Multiple-Input-Multiple-Output (DL/UL MU-MIMO), spatial reuse techniques like BSS coloring, and efficient use of channel resources like DL/UL Orthogonal Frequency Division Multiple Access (OFDMA). Therefore, it becomes necessary to provide reliable simulation tools able to assess the performance and behavior of next-generation WLANs in multiple scenarios/cases, especially in high-density deployments.
	
	In this paper, we present Komondor,\footnote{All of the source code of Komondor, under the GNU General Public License v3.0., is open, and potential contributors are encouraged to participate. The repository can be found at \url{https://github.com/wn-upf/Komondor}} an open source, event-driven simulator based on the CompC++ COST library \cite{chen2002reusing}. Komondor is focused on fulfilling the need for assessing the novel features introduced in recent and future amendments, which may be endowed with applications driven by machine learning (ML). The motivation for developing and building the presented wireless network simulator is motivated by: 
	\begin{enumerate}[label=\roman*)]
		\item The lack of analytical models for capturing next-generation techniques in spatially distributed and/or high-density deployments.
		\item The lack of next-generation WLAN-oriented simulators.
		\item The complexity of extending simulators comprising an exhaustive implementation of the physical (PHY) layer.
		\item The large (or intractable) execution time required by other simulators to simulate high-density deployments.
		\item The need for conveniently incorporating ML-based agents in the simulation tool.
	\end{enumerate}
	In short, Komondor is designed to efficiently implement new functionalities by relying on flexible and simplified PHY layer dependencies, to be faster than most off-the-shelf simulators, and to provide reliable simulations and a gentle learning curve to new users. 
	
	\section{Wireless network simulators}
	\label{section:related_work}
	
	Wireless network simulators can be categorized into continuous-time and discrete-event. On the one hand, continuous-time simulators continuously keep track of the system dynamics by dividing the simulation time into very small periods of time. On the other hand, in discrete-event simulators, events are used to characterize changes in the system. Accordingly, for the latter, events are ordered in time and normally allow running faster simulations than continuous-time simulators. In addition, discrete-event simulators allow tracing events with higher precision.
	
	From the family of discrete-event driven network simulators, only a few ones are publicly available. OMNET++~\cite{varga2008overview} is a component-based C++ simulation library that is not open-source and is used for modeling communication networks and distributed multiprocessor systems. OPNET is another commercial simulator that allows the integration of external components. NetSim~\cite{rathi1990new} was conceived to provide an accurate simulation model oriented to the world wide web. To that purpose, the simulator was written in Java, which compromises simulation time with programming flexibility. When it comes to open source network simulators, a MATLAB-based link-level simulator was presented in~\cite{milos2016link} for supporting the IEEE 802.11g/n/ac/ah/af technologies. The ns-2 simulator~\cite{issariyakul2012introduction} is another network simulator known for its accuracy and the integration with the network animator. Finally, the ns-3, which was introduced in 2006 to replace the ns-2, presents significant advantages over the ns-2 due to its detailed simulation features, becoming very popular among the research community~\cite{riley2010ns}. Table~\ref{table:simulators_comparison} highlights in a nutshell the most important characteristics of the overviewed network simulators and Komondor.
	
	\begin{table}[t!]
		\footnotesize
		\caption{Comparison of wireless network simulators.}
		\label{table:simulators_comparison}
		\resizebox{\columnwidth}{!}{\begin{tabular}{@{}ccccccc@{}}
				\toprule
				\textbf{Simulator} & \begin{tabular}[c]{@{}c@{}}\textbf{Open-}\\ \textbf{source}\end{tabular} & \begin{tabular}[c]{@{}c@{}}\textbf{Source}\\ \textbf{lang.}\end{tabular} & \textbf{Complexity} & \begin{tabular}[c]{@{}c@{}}\textbf{GUI}\end{tabular} & \begin{tabular}[c]{@{}c@{}}\textbf{11ax}\\ \textbf{features}\end{tabular} & \begin{tabular}[c]{@{}c@{}}\textbf{ML/based}\\ \textbf{module}\end{tabular} \\ \midrule
				ns-3      & Yes         & C++                                                       & High       & No\footnotemark[1]                                                        & Partial                                                 & No\footnotemark[2]                                                         \\
				ns-2      & Yes         & C++/OTcl                                                  & Low        & No\footnotemark[1]                                                      & No                                                      & No                                                        \\
				OMNET++   & No          & C/C++                                                     & Medium     & Yes                                                         & No                                                      & No                                                        \\
				OPNET     & No          & C++                                                       & Medium     & Yes                                                         & No                                                      & No                                                        \\
				NetSim    & No          & Java                                                      & Low        & Yes                                                         & No                                                      & No                                                        \\
				\textbf{Komondor}  & \textbf{Yes}         & \textbf{C/C++}                                                     & \textbf{Low}        & \textbf{No}                                                          & \textbf{Partial}                                                 & \textbf{Yes}                                                       \\ \bottomrule
		\end{tabular}}
	\end{table}
	\footnotetext[2]{\label{note1}Although ns-2 and ns-3 do not provide a default graphical animation tool, there are tools supporting live animation, e.g., PyViz or NetAnim for ns-3 and NAM for ns-2.}
    \footnotetext[3]{blueAn integration with OpenAI Gym has been recently provided to ns-3 \cite{gawlowicz2018ns3}, but the ML-based operation is not part of the simulator.}  
	
	Among the family of overviewed discrete-event simulators, we highlight the ns-3 open-source simulator due to its popularity and use it as a baseline for comparing against Komondor. Despite the plethora of features that are supported in ns-3, it has some inherent limitations, such as the high complexity for developing new features/models as an extension of the simulator core. In particular, compatibility with the already existing/supported models is required and must be carefully ensured. For example, beamforming for previous mature amendments (i.e. IEEE 802.11n/ac) is not available yet, owing to the effort required to integrate it. Moreover, the integration of new features strongly depends on the willingness of the community to contribute to the development.
	
	With respect to the IEEE 802.11ax operation -- rates and support for information elements are being developed -- the implementation is mostly based on the Draft 1.0~\cite{stacey2016proposed}. Such a draft dates from 2016 and does not include most of the core IEEE 802.11ax functionalities. At the time of submitting this paper, only the Single-User Protocol Data Unit (SU PPDU) and MIMO with up to four antennas are supported in ns-3, whereas OFDMA and MU-MIMO are not supported in the official distribution~\cite{ns3documentation}.
	
	Apart from the official resources, we find few ns-3 works publicly available that support IEEE 802.11ax features, which may (or may not) be integrated into future releases. For example, we highlight the works with regard to the OFDMA that have been carried out by Getachew Redieteab et al. (based on the IEEE 802.11ax specification framework document~\cite{stacey2016specification}) and Cisco~\cite{cisco2017simulator}. However, none of these works completely follow the latest developments in the IEEE 802.11ax standard and have not been validated through extensive simulations and testbed results, as had previously occurred with the OFDM~\cite{pei2010validation}. In addition to OFDMA, the spatial reuse operation (i.e., BSS Color~\cite{7794832_color}) is under active development, whereas extensions of the capture effect have been applied to ns-3 to follow the IEEE 802.11ax guidelines and studied in~\cite{selinis2017exploiting} and in a testbed~\cite{8433688_KhorovCE}.
	
	\section{Komondor Design Principles}
	\label{section:system_model}
	
	\subsection{Architecture}
	\label{subsection:architecture}
\begin{figure}[h!]
		\centering
		\includegraphics[width=0.7\columnwidth]{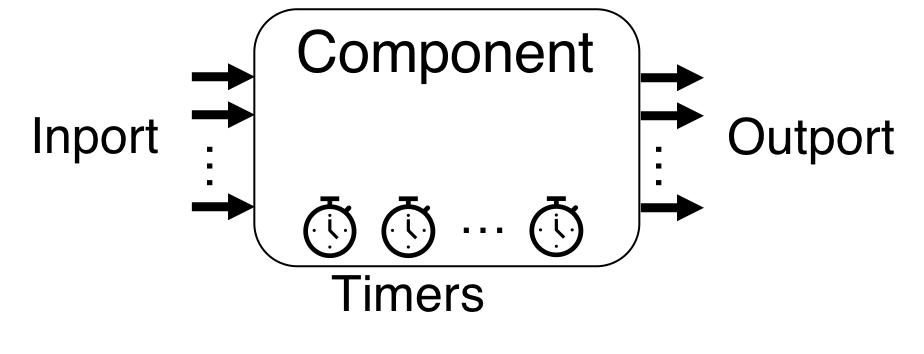}
		\caption{COST component. While inports and outports allow to directly communicate with other components, timers trigger events specific to the component.}
		\label{fig:cost}
	\end{figure}

	Komondor aims to realistically capture the operation of WLANs. Henceforth, it reproduces actual transmissions on a per-packet basis. To that purpose, Komondor is based on the COST library, which allows building interactions between components (e.g., wireless nodes, buffers, packets) through synchronous and asynchronous events. While the former are messages explicitly exchanged between components through input/output ports, the latter are based on timers.
	In practice, components perform a set of operations until a significant event occurs. For instance, a node that is decreasing its backoff may freeze it when an overlapping node occupies the channel. The beginning and end of such a transmission are examples of significant events, whereas decreasing the backoff counter is not. Nevertheless, events may be triggered by different timers. In the previous example, a node's transmission begins once the backoff timer terminates (i.e., the backoff timer triggers the beginning of the transmission), while the end of the transmission is triggered by the packet transmission timer. Fig.~\ref{fig:cost} shows the schematic of a COST component, which is composed of inports, outports, and a set of timers. 
	
	\subsection{IEEE 802.11 Features}
	\label{subsection:features}
    Komondor entails a long-term project in which several contributors are involved. That is, the simulator is continuously evolving to include novel techniques and generally improve performance. The current version of Komondor (v2.0) includes the following fully tested IEEE 802.11ax features:
	\begin{itemize}
		\item \textbf{Distributed coordination function (DCF)}: the Carrier Sense Multiple Access with Collision Avoidance (CSMA/CA) captures the basic Wi-Fi operation for accessing the channel. Moreover, Contention Window (CW) adaptation is considered.
		\item \textbf{Buffering and packet aggregation}: several traffic generator models are implemented in Komondor such as deterministic, Poisson or full-buffer. Besides, multiple media access control protocol data unit (MPDU) can be aggregated into the same PLCP protocol data unit (PPDU) in order to reduce the generated communication overheads.
		\item \textbf{Dynamic channel bonding (DCB)}: multiple channel widths can be selected during transmissions by implementing DCB policies in order to maximize the spectrum efficiency. Some of these policies were already evaluated in~\cite{barrachina2019dynamic},\cite{barrachina2019overlap}.
		\item\textbf{Modulation coding scheme (MCS) selection}: the MCS is negotiated between any transmitter-receiver pair according to the Signal-to-Interference-and-Noise Ratio (SINR), thus supporting multiple transmission rates.
		\item \textbf{Ready-to-send/Clear-to-send (RTS/CTS) and Network Allocation Vector (NAV)}: virtual carrier sensing is implemented in order to minimize the number of collisions by hidden-nodes.
	\end{itemize}
	
	Future development stages are under progress including other features such as OFDMA, MU-MIMO transmissions, beamforming, spatial reuse, and ML-based configuration.
	
	\subsection{Execution Flowchart}
	\label{subsection:flowchart}
\begin{figure}[t]
		\centering
		\includegraphics[width=0.9\columnwidth]{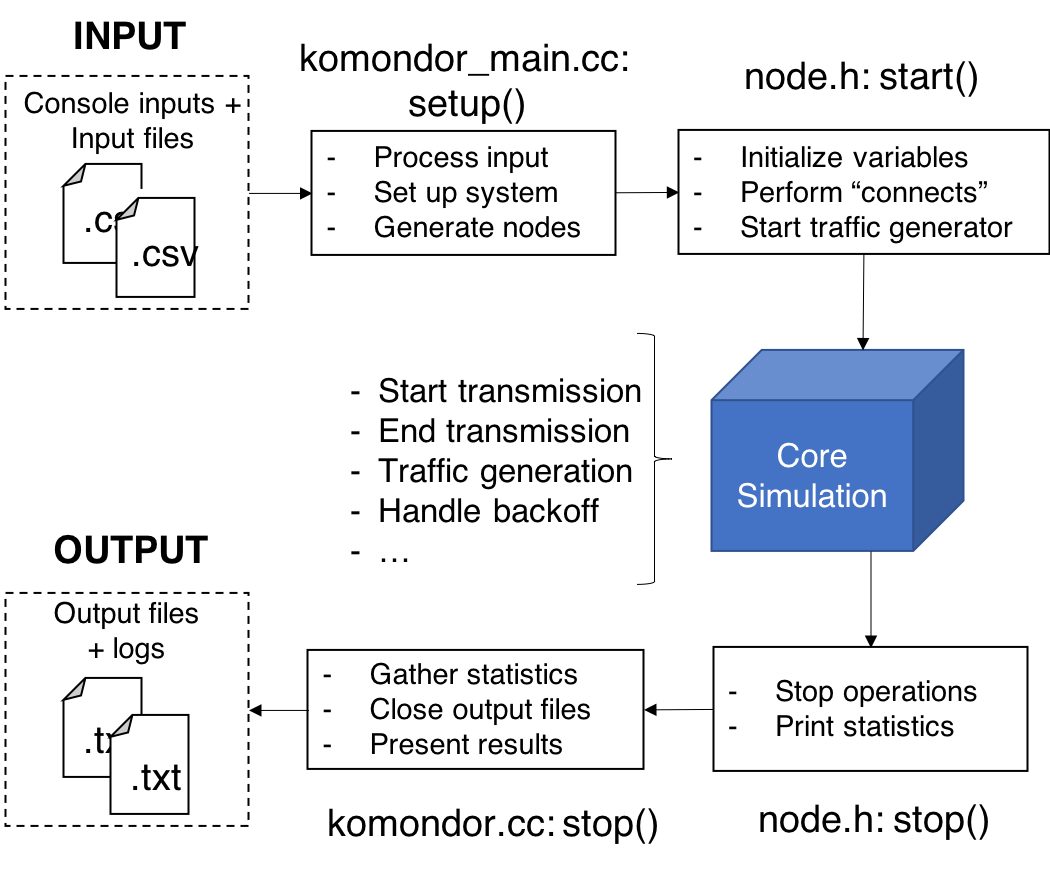}
		\caption{Komondor execution flowchart.}
		\label{fig:komondor_flowchart}
	\end{figure}

	Komondor is composed of several modules that allow performing simulations with a high degree of freedom. Fig.~\ref{fig:komondor_flowchart} summarizes the operational mode of Komondor from a user's point of view. A more detailed user's guide providing a quick-start and guided execution examples is available in the Komondor's Github repository.
	
	\subsubsection{Input and Setup/Start}
	as for the execution console command for starting Komondor simulations, arguments are designed in a simple and efficient way. Examples of console arguments are the file names of the inputs, the activation flags of the logs, the simulation time and the random seed. In addition, input files (in CSV format) are used to define the environment and have been conceived in a way that the user can easily modify important simulation parameters such as the traffic load, the path-loss model, or the data packet size.
	Once the environment is generated and nodes are initialized, traffic is exchanged between nodes until the simulation time runs out.
	
	\subsubsection{Stop and Output}
	when the simulation finishes, the closing is handled and statistics are gathered. Then, extensive and detailed performance statistics are per default provided by Komondor (e.g., throughput, delay, spectrum utilization, or collisions). Moreover, the user can efficiently include as much as metrics as desired.
	
	\begin{figure*}[t]
		\centering
		\includegraphics[width=0.98\textwidth]{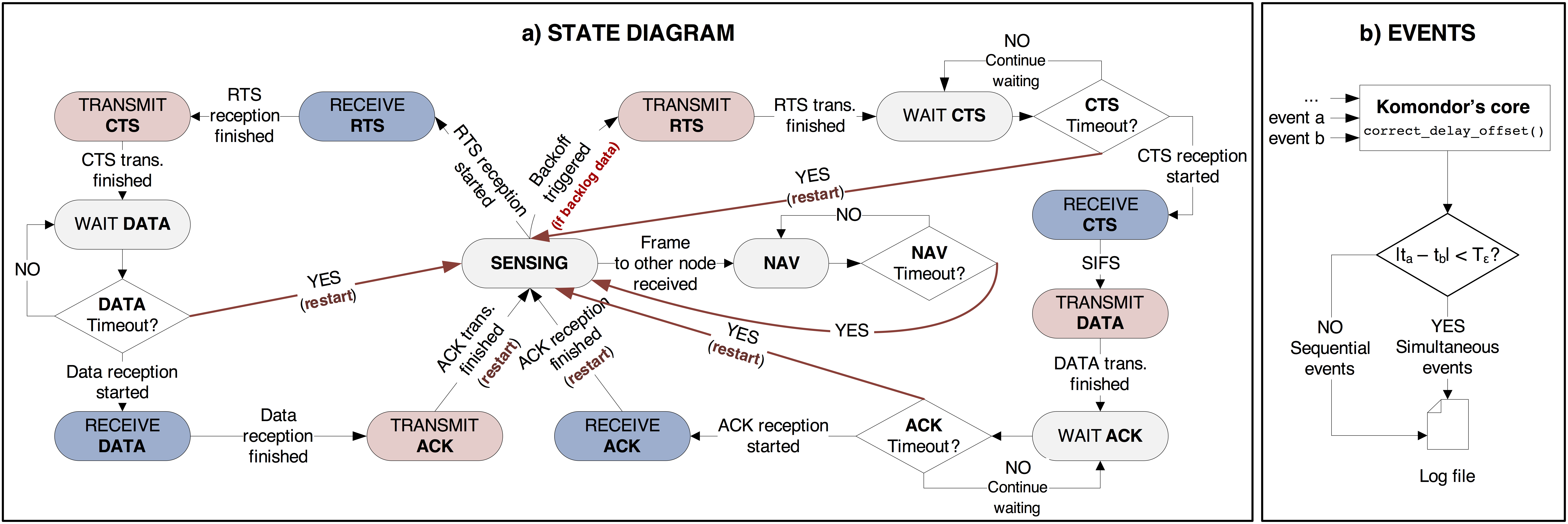}
		\caption{Komondor's state diagram and events. a) States are reachable by different transitions. b) Simultaneous events are properly processed through delay offset correction.}
		\label{fig:state_diagram}
	\end{figure*}
	
	\subsection{States and events}
	\label{subsection:state_diagram}
	The Komondor's core operation is based on states, which represent the status (or situation) in which a node can be involved. A state diagram summarizing both states and transitions is shown in Fig.~\ref{fig:state_diagram}. Roughly, a given node starts in the SENSING state, where multiple events can occur (e.g., a new packet is buffered or a new transmission is detected). Then, according to the noticed event, the node transits to the corresponding state. 
	
	\subsubsection{States}
	we depict below each state and how a node must behave in front of new events.
	
	\begin{itemize}
		
		\item \textbf{SENSING}: a node senses the channel with two main purposes. First, to follow the CSMA/CA operation to gain access to the channel (in case there is backlogged data in the buffer/s). Second, to wait for incoming transmissions, so that either carrier sensing or receiving procedures are held. In case of being immersed in a backoff procedure, a node detecting a ``new transmission'' event would sense the power received in its primary channel, and assess whether to freeze the backoff countdown. Similarly, whenever an ``end transmission'' event occurs, the channel is sensed in order to determine whether the backoff counter can be resumed or remain paused.
		
		\item \textbf{TRANSMIT}: a transmitter node is currently transmitting a packet. No matter what events may occur, during the packet transmission, the node blocks its receiver capabilities and remains in the same state until the transmission is finished.
		
		\item \textbf{RECEIVE}: when a node is receiving and decoding an incoming packet, it will behave in front of a new event according to its implication in the channel of interest. Of especial importance are those new transmission events triggered by other nodes that have gained access to the medium. Specifically, if a new transmission generates enough interference, the ongoing reception will be discarded, thus leading to a packet loss. 
		
		\item \textbf{WAIT states}: these states allow modeling the situations where a node that transmitted a packet is expecting for the corresponding response. Namely, after transmitting RTS, CTS or DATA packets, the transmitter will wait for the corresponding CTS, DATA or ACK/BACK packets, respectively. If the response packet is not received before the corresponding timeout is triggered, the transmitter assumes that either the transmitted packet or the response packet is lost and resets to SENSING state. Wait states are particularly useful to detect packet losses when anomalies in the network (e.g., hidden terminal problem) occur.
		
		\item \textbf{NAV}: when a node enters in NAV state due to the successful reception of a frame addressed to a different destination, it sets a NAV timer and keeps listening to its primary channel. If a new frame is successfully received during the NAV, the timer is updated, provided that the new NAV time is larger than the current remaining time.
	\end{itemize}
	
	\subsubsection{Events}
	each time a node performs an action that can affect the system (e.g., it starts transmitting a frame), an event is announced. Events in Komondor are lined up on the time axis and handled by the core entity. Events management is similar in ns-3. However, the latter exhibits a significant limitation, since events that are scheduled at the exact same time can be executed in any order. Such a development feature may lead to unpredictable results and is incompatible with real-world situations in which events can occur simultaneously. Some inconsistencies may occur in case that the execution order affects multiple simultaneous events (e.g., two packets arriving at the exact same time). To solve this, Komondor, which is also a discrete-event simulator, employs temporal variables to compare the exact timestamps at which two or more events were generated. As a result, Komondor is able to successfully simulate the behavior of simultaneous events while keeping the logic of the states.
	
	\subsection{Developing new modules}
	
	Komondor has been conceived to be easily modified and extended. In particular, several modules have been provided to represent different simulation capabilities (e.g., propagation, channel access or traffic generation). Accordingly, Komondor can be potentially extended to support the operation of other IEEE 802.11 amendments such as 11n, 11ac, 11ad or 11ay. In addition, ML-based modules can also be introduced. A complete manual can be found at the Komondor's repository.
	
	\section{Validation}
	\label{section:validations}
	In this Section, we validate the operation of Komondor and show its potential for dealing with high-density scenarios. In particular, we show the reliability of the simulator, despite its reduced complexity of the PHY.\footnote[4]{For instance, channel effects are assumed to remain static during the whole transmission of a given frame, and the propagation delay is considered to be negligible.} The validation of the Komondor's operation is done through a set of illustrative scenarios, and our results are compared with the ones obtained with ns-3.\footnote[5]{Details on the ns-3 implementation used in the simulations presented throughout this paper can be found at \url{https://github.com/wn-upf/Komondor/tree/master/Documentation/Validation/ns-3}. For instance, this implementation includes the 11ax residential scenario propagation loss~\cite{pathloss11ax} and has a PLCP training duration updated according to the 11ax amendment~\cite{tgax2017draft}.} In addition to ns-3, a mutual validation is performed with the Continuous Time Markov Networks (CTMNs) modeling introduced in~\cite{bellalta2014throughput}, and which is extended for spatially distributed networks in the Spatial-Flexible Continuous Time Markov Network (SFCTMN) framework~\cite{barrachina2019dynamic}. As for high-density scenarios, we make use of the Bianchi's DCF analytical model~\cite{bianchi2000performance} to validate the results in fully-overlapping deployments, where all the nodes are within the carrier sense of the others. The results shown in the following subsections were obtained according to the parameters defined in Table~\ref{table:parameters}. The duration of the RTS, CTS and data frame is computed as follows:
	
	\begin{flalign*}
	T_\text{RTS} = T_{\text{PHY-leg}} + \ceil*{\frac{L_{SF} + L_\text{RTS}}{L_{s,l}}} \sigma_\text{leg}  \text{,} \\
	T_\text{CTS} = T_{\text{PHY-leg}} + \ceil*{\frac{ L_{SF} + L_\text{CTS}}{L_{s,l}}} \sigma_\text{leg} \text{,} \\
	T_\text{D} = T_{\text{HE-SU}} + \ceil*{\frac{L_{\text{SF}} + L_{\text{MH}} + N_{\text{agg}} L_\text{D}}{L_{s,l}}} \sigma \text{.}
	\end{flalign*}
	
	Note that full-buffer traffic is assumed in all the scenarios throughout this work for comparative purposes. Moreover, we have considered the residential path-loss model recommended in the IEEE 802.11ax \cite{pathloss11ax}, which inflicts high losses due to its large number of obstacles (e.g., walls).
	
	\begin{table}[h]
		\caption{Parameters considered in the presented scenarios.}
		\label{table:parameters}
		\centering
		\begin{tabularx}{\columnwidth}{sbs}
			\toprule
			\textbf{Parameter}     & \textbf{Description}              & \textbf{Value} \\ 
			\midrule
			$f_\text{c}$ & Central frequency           & 5 GHz  \\
			$|c|$ & Basic channel bandwidth          & 20 MHz \\
			MCS						& 11ax MCS index							& 0-11		\\
			$G_\text{tx}$         & Transmitting gain                 & 0 dB           \\ 
			$G_\text{rx}$         & Reception gain                    & 0 dB           \\ 
			$\text{PL}(d)$		& Path loss (Residential scenario)	&  see~\cite{pathloss11ax}		\\
			$N$                      & Background noise level            & -95 dBm        \\
			$\sigma_\text{leg}$      & Legacy OFDM symbol duration     & 4 \textmu s           \\
			$\sigma$      & OFDM symbol duration (GI-32)     & 16 \textmu s           \\
			$N_{sc}$      & Number of subcarriers (20 MHz)     & 234          \\
			$N_{ss}$      & Number of spatial streams     & 1          \\
			\midrule
			$T_\text{e}$       & Empty slot duration                     & 9 \textmu s          \\
			$T_\text{SIFS}$                   & SIFS duration                     & 16 \textmu s      \\ 
			$T_\text{DIFS}$                   & DIFS duration                     & 34 \textmu s      \\ 
			$T_\text{PIFS}$                   & PIFS duration                     & 25 \textmu s      \\
			$T_\text{PHY-leg}$      & Legacy preamble duration   & 20 \textmu s           \\
			$T_\text{HE-SU}$      & HE single-user field duration       & 100 \textmu s \\
			$T_\text{ACK}$       & ACK duration           & 28 \textmu s        \\ 
			$T_\text{BACK}$       & Block ACK duration         & 32 \textmu s       \\
			$T_\text{PPDU}^{\text{max}}$       & Max. PPDU duration     &  5484 \textmu s       \\
			$L_{s,l}$       &  Size OFDM symbol (legacy)         & 24 bits     \\ 
			$L_\text{D}$       & Data packet size           & 11728 bits     \\ 
			$N_{\text{agg}}$       & No. of frames in an A-MPDU & 1, 64             \\
			$L_\text{RTS}$        & Length of an RTS packet           & 160 bits       \\ 
			$L_\text{CTS}$        & Length of a CTS packet            & 112 bits       \\ 
			$L_\text{SF}$      & Length of service field       & 16 bits           \\ 
			$L_\text{MH}$      & Length of MAC header     & 320 bits           \\
			$\text{CW}$ & Contention window (fixed)          & 15           \\
			\bottomrule
		\end{tabularx}
	\end{table}
	
	\subsection{Analyzing toy Scenarios}
	
	Komondor has been conceived as a friendly and ready-to-use wireless network simulator that can be used by researches and teachers to study fundamental networking issues. In particular, scenarios and environment configurations can be conveniently modified through structured input files. The scenarios proposed in this Section are a clear example of toy scenarios where different networking concepts such as \textit{flow starvation} or \textit{additive interference} take place. Furthermore, a given user can easily analyze WLAN scenarios through the implemented logs generation system and statistics reporting. Accordingly, particular phenomena in the PHY and medium access control (MAC) layers can be tracked (e.g., channel contention, packet collisions, physical carrier sensing, energy detection, or buffer dynamics).
	
	\subsection{Basic Operation}
	\label{section:simple_scenarios}
	
	\begin{figure}[t]
		\centering		
		\includegraphics[width=0.8\columnwidth]{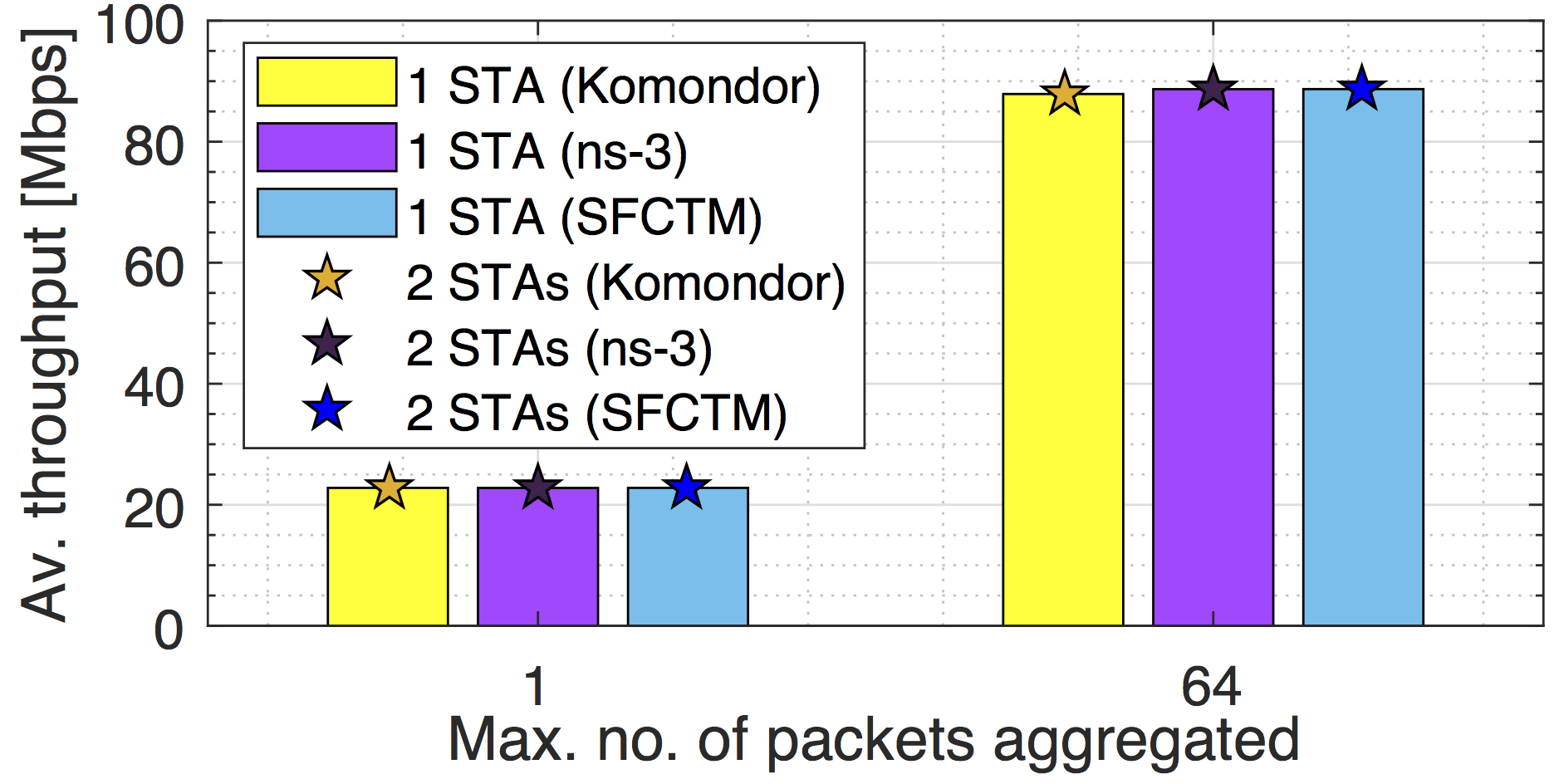}
		\caption{Average throughput experienced by the WLAN of \textit{Scenario 1}, for $N_{\text{agg}} = 1$ and $N_{\text{agg}} = 64$. Results obtained from each simulation tool are shown.}
		\label{fig:results_simple_scenarios}
	\end{figure}
	
	\begin{figure*}[ht!]
		\begin{subfigure}{0.225\textwidth}
			\includegraphics[width=\linewidth]{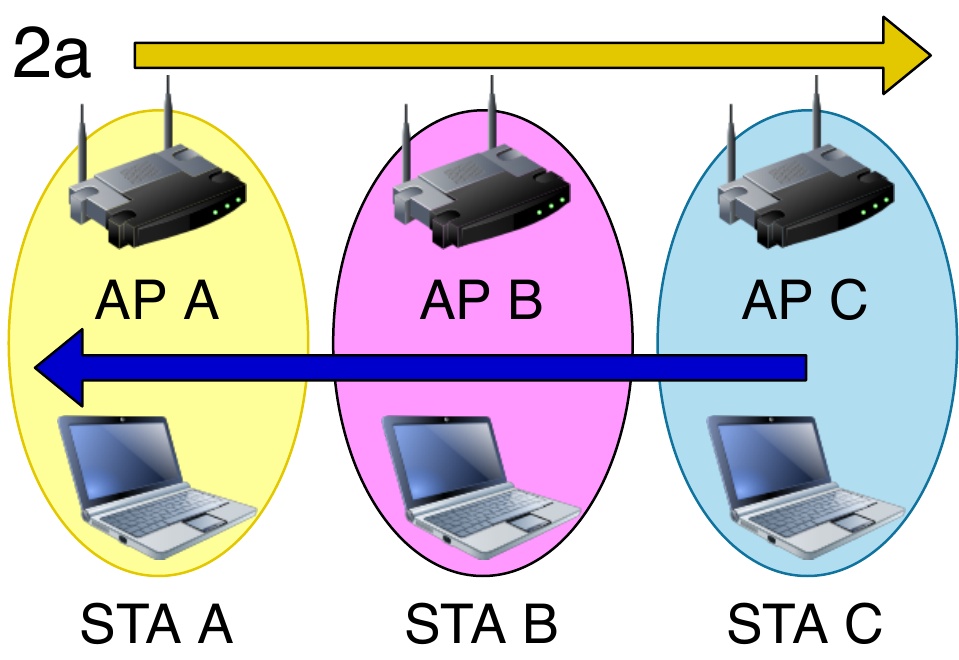}
			\caption{Topology of \textit{Scenario 2a.}} \label{fig:x_a}
		\end{subfigure}
		\hspace{0.1mm}
		\begin{subfigure}{0.24\textwidth}
			\begin{tikzpicture}[<->,>=stealth',shorten >=1pt,auto,node distance=1.8cm,
			semithick]
			\tikzstyle{every state}=[fill=white,draw=black,thick,text=black,scale=0.72]
			\node[state, label=above left:2a]    (S1)                    {$\emptyset$};
			\node[state]    (S3)[right of = S1, xshift=-0.3cm]		{$\text{B}$};
			\node[state]    (S2)[above of = S3, yshift=-0.3cm]		{$\text{A}$};
			\node[state]    (S4)[below of = S3, yshift=0.3cm]		{$\text{C}$};
			\node[state, draw=white]    (S99)[right of = S3, xshift=-0.3cm]		{};
			\node[state, draw=white]    (S999)[left of = S1, xshift=0.3cm]		{};
			
			\path
			(S1) edge[bend left] (S2)
			(S1) edge (S3)
			(S1) edge[bend right] (S4);
			
			\end{tikzpicture}
			\caption{CTMN of \textit{Scenario 2a.}} \label{fig:c_a}
		\end{subfigure}
		\hspace*{\fill}
		\begin{subfigure}{0.225\textwidth}
			\includegraphics[width=\linewidth]{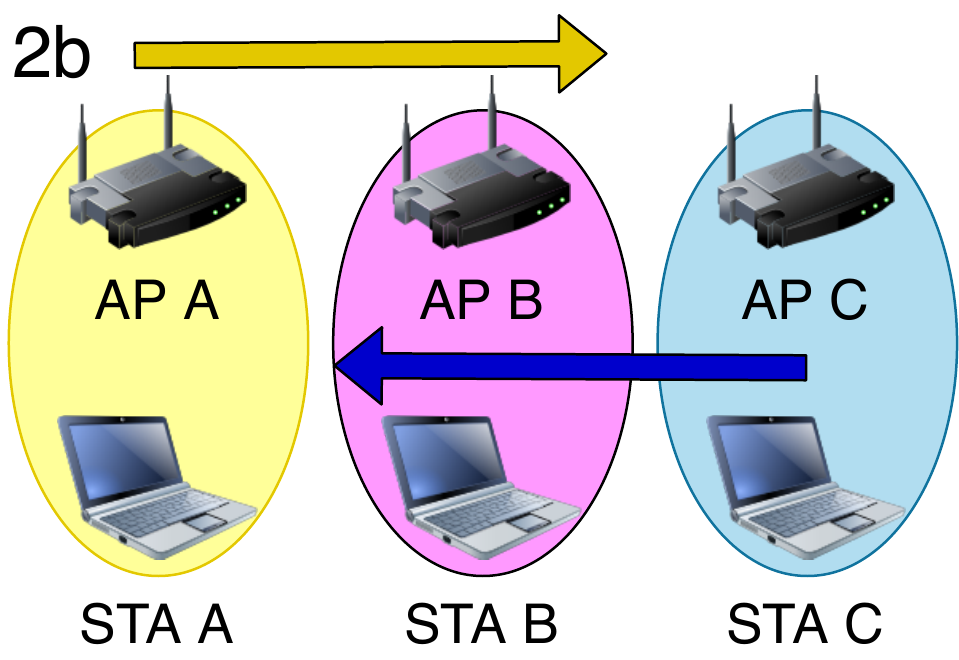}
			\caption{Topology of \textit{Scenario 2b}.} \label{fig:x_b}
		\end{subfigure}
		\hspace{0.1mm}
		\begin{subfigure}{0.24\textwidth}
			\begin{tikzpicture}[<->,>=stealth',shorten >=1pt,auto,node distance=1.8cm,
			semithick]
			\tikzstyle{every state}=[fill=white,draw=black,thick,text=black,scale=0.72]
			
			\node[state, label=above left:2b]    (S1)                    {$\emptyset$};
			\node[state]    (S3)[right of = S1, xshift=-0.3cm]		{$\text{B}$};
			\node[state]    (S2)[above of = S3, yshift=-0.3cm]		{$\text{A}$};
			\node[state]    (S4)[below of = S3, yshift=0.3cm]		{$\text{C}$};
			\node[state]    (S6)[right of = S3, xshift=-0.3cm]		{$\text{A}\text{C}$};
			\node[state, draw=white]    (S99)[right of = S6, xshift=-0.885cm]		{};
			\node[state, draw=white]    (S999)[left of = S1, xshift=0.885cm]		{};
			
			\path
			
			(S1) edge[bend left] (S2)
			(S1) edge (S3)
			(S1) edge[bend right] (S4)
			(S2) edge (S6)
			(S4) edge (S6);
			\end{tikzpicture}
			\caption{CTMN of \textit{Scenario 2b}.} \label{fig:c_b}
		\end{subfigure}
		
		\vspace*{2mm}
		\begin{subfigure}{0.225\textwidth}
			\includegraphics[width=\linewidth]{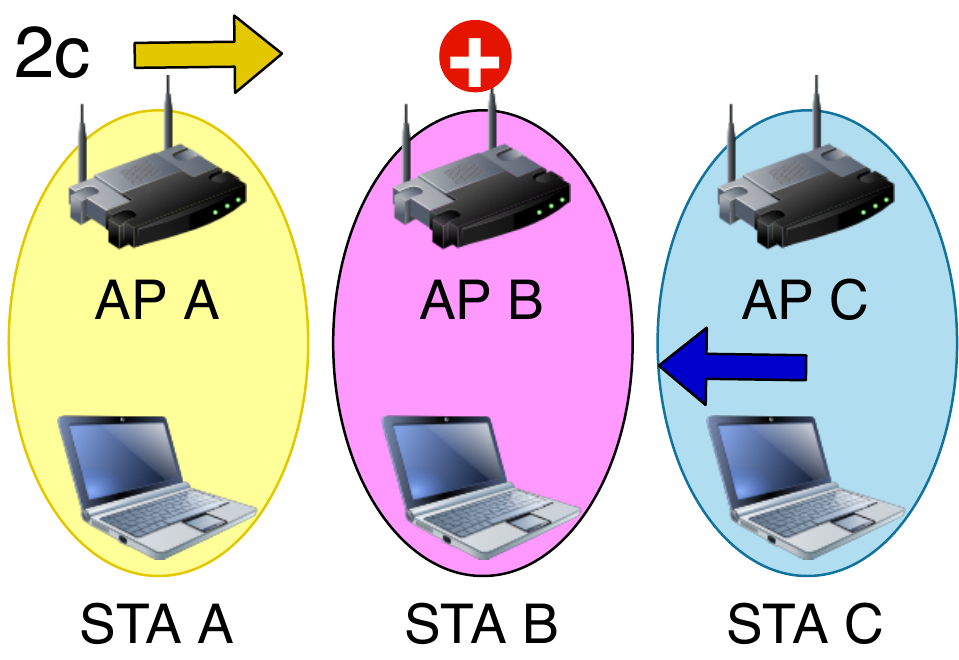}
			\caption{Topology of \textit{Scenario 2c}.} \label{fig:x_c}
		\end{subfigure}
		\hspace*{\fill}
		\begin{subfigure}{0.24\textwidth}
			\begin{tikzpicture}[<->,>=stealth',shorten >=1pt,auto,node distance=1.8cm,
			semithick]
			\tikzstyle{every state}=[fill=white,draw=black,thick,text=black,scale=0.72]
			
			\node[state, label={[yshift=0.4cm]2c}]    (S1)                    {$\emptyset$};
			
			\node[state]    (S3)[right of = S1, xshift=-0.3cm]		{$\text{B}$};
			
			\node[state]    (S2)[above of = S3, yshift=-0.3cm]		{$\text{A}$};
			
			\node[state]    (S4)[below of = S3, yshift=0.3cm]		{$\text{C}$};
			
			\node[state]    (S6)[right of = S3, xshift=-0.3cm]		{$\text{A}\text{C}$};
			
			\node[state]    (S5)[above of = S6, yshift=-0.3cm]		{$\text{A}\text{B}$};
			
			\node[state]    (S7)[below of = S6, yshift=0.3cm]		{$\text{B}\text{C}$};
			
			\node[state]    (S8)[right of = S6, xshift=-0.3cm]		{$\text{A}\text{B}\text{C}$};
			
			\path
			
			(S1) edge[bend left] (S2)
			
			(S1) edge (S3)
			
			(S1) edge[bend right] (S4)
			
			(S2) edge (S5)
			
			(S2) edge (S6)
			
			(S3) edge (S5)
			
			(S3) edge (S7)
			
			(S4) edge (S6)
			
			(S4) edge (S7)
			
			(S5) edge (S8)
			
			(S6) edge[<-, red] (S8)
			
			(S7) edge (S8);
			
			\end{tikzpicture}
			\caption{CTMN of \textit{Scenario 2c}.} \label{fig:c_c}
		\end{subfigure}
		\hspace*{\fill}
		\begin{subfigure}{0.225\textwidth}
			\includegraphics[width=\linewidth]{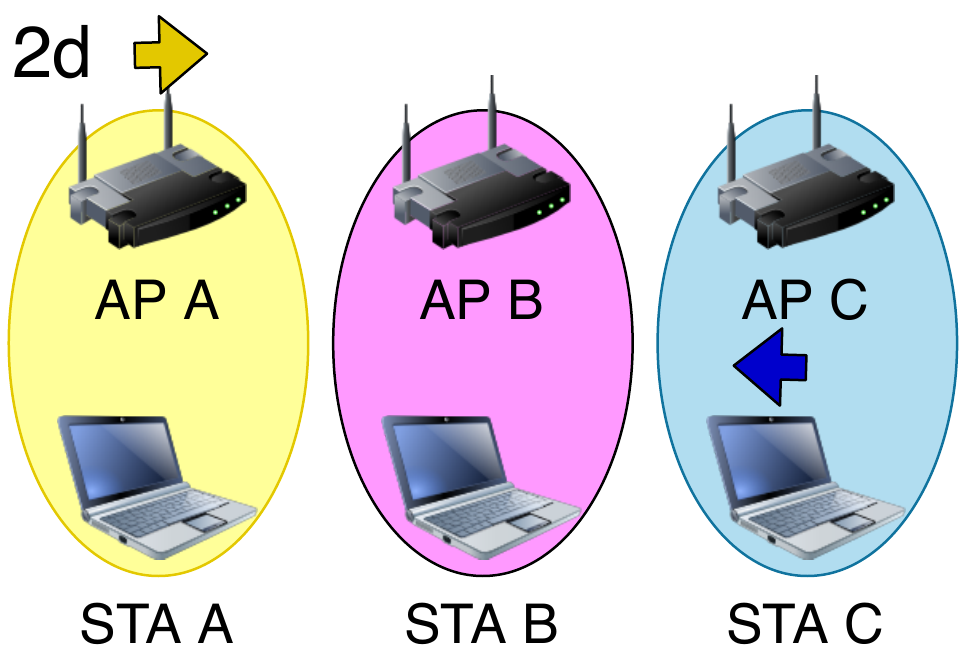}
			\caption{Topology of \textit{Scenario 2d}.} \label{fig:x_d}
		\end{subfigure}
		\hspace*{\fill}
		\begin{subfigure}{0.24\textwidth}
			\begin{tikzpicture}[<->,>=stealth',shorten >=1pt,auto,node distance=1.8cm,
			semithick]
			\tikzstyle{every state}=[fill=white,draw=black,thick,text=black,scale=0.72]
			
			\node[state, label={[yshift=0.4cm]2d}]    (S1)                    {$\emptyset$};
			
			\node[state]    (S3)[right of = S1, xshift=-0.3cm]		{$\text{B}$};
			
			\node[state]    (S2)[above of = S3, yshift=-0.3cm]		{$\text{A}$};
			
			\node[state]    (S4)[below of = S3, yshift=0.3cm]		{$\text{C}$};
			
			\node[state]    (S6)[right of = S3, xshift=-0.3cm]		{$\text{A}\text{C}$};
			
			\node[state]    (S5)[above of = S6, yshift=-0.3cm]		{$\text{A}\text{B}$};
			
			\node[state]    (S7)[below of = S6, yshift=0.3cm]		{$\text{B}\text{C}$};
			
			\node[state]    (S8)[right of = S6, xshift=-0.3cm]		{$\text{A}\text{B}\text{C}$};
			
			\path
			
			(S1) edge[bend left] (S2)
			
			(S1) edge (S3)
			
			(S1) edge[bend right] (S4)
			
			(S2) edge (S5)
			
			(S2) edge (S6)
			
			(S3) edge (S5)
			
			(S3) edge (S7)
			
			(S4) edge (S6)
			
			(S4) edge (S7)
			
			(S5) edge (S8)
			
			(S6) edge (S8)
			
			(S7) edge (S8);
			\end{tikzpicture}
			\caption{CTMN of \textit{Scenario 2d}.} \label{fig:c_d}
		\end{subfigure}
		\caption{Topologies and corresponding CTMNs of scenarios 2a-2d. The yellow and blue arrows represent the area of interference from transmitters in WLANs A and C, respectively, whereby medium contention is forced.}
		\label{fig:complex_scenarios}
	\end{figure*}
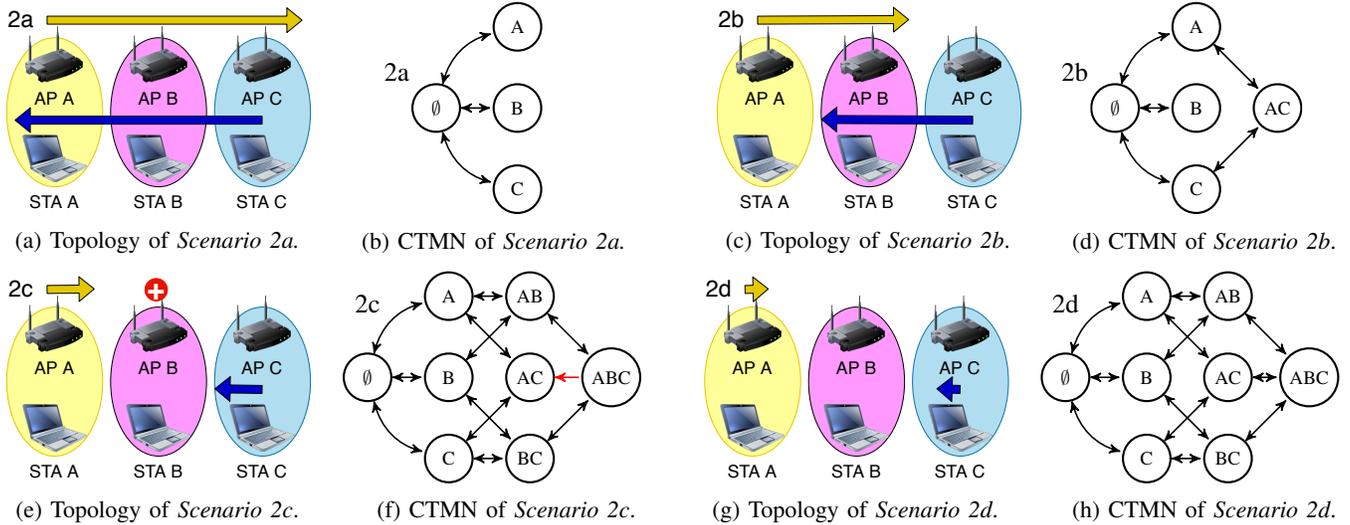
	
	We first aim to validate the basic IEEE 802.11 operation of the DCF implemented in Komondor when RTS/CTS is applied. For that, we consider a single Access Point (AP) scenario (we name it \textit{Scenario 1}) with one and two stations (STAs), where full-buffer downlink traffic is held. The two-STAs case allows us to assess the proper behavior of Komondor in presence of multiple STAs. To validate this scenario, we compare the Komondor results with the ones provided by ns-3 and the SFCTMN framework. Fig.~\ref{fig:results_simple_scenarios} shows the simulation results obtained from each tool, for packet aggregation ($N_{\text{agg}} = 64$) and no-aggregation ($N_{\text{agg}} = 1$). We note that the average throughput obtained by each simulation tool is almost identical, either for packet aggregation or not. In addition, having multiple STAs leads to the same result as for a single one since the destination STA is picked at random in every transmission.
	
	\subsection{Complex inter-WLAN interactions}
	\label{section:complex_scenarios}
	In order to validate the behavior of Komondor in front of more complex inter-WLAN interactions, we now focus on the three-WLANs scenarios shown in Fig.~\ref{fig:complex_scenarios}. We name them \textit{Scenario 2a-2d}. The interactions occurring in such scenarios are illustrated through CTMNs, where states\footnote[6]{Note that CTMN states are not related by any means to Komondor states.} represent the WLANs that are currently transmitting. Note that each of these scenarios reflects different situations that are of particular interest since they generalize different well-known phenomena in wireless networks:
	\begin{itemize}
		
		\item \textbf{Fully overlapping (Fig.~\ref{fig:x_a})}: all the nodes cause contention to all the others when transmitting. For that, the distance between consecutive APs and between AP and STA of the same WLAN is set to $d_{\text{AP,AP}} = d_{\text{AP,STA}} = 2$ m, respectively.
		
		\item \textbf{Flow starvation (Fig.~\ref{fig:x_b})}: contention is triggered in a pair-wise manner, so that $\text{WLAN}_\text{A}$ and $\text{WLAN}_\text{C}$ do not interfere each other. For that, the distance is set to $d_{\text{AP,AP}} = 4$ m and $d_{\text{AP,STA}} = 2$ m. Note that this case could be also extended to show a hidden node effect if $\text{AP}_\text{A}$ or $\text{AP}_\text{C}$ were intended to transmit to a STA located at the location of $\text{AP}_\text{B}$.
		
		\item \textbf{Potential overlap (Fig.~\ref{fig:x_c})}:
		contention only occurs at $\text{WLAN}_\text{B}$ when both $\text{WLAN}_\text{A}$ and $\text{WLAN}_\text{C}$ transmit concurrently. Otherwise, the channel is sensed as free. Note that, in this case, packets are successfully transmitted in $\text{WLAN}_\text{B}$ whenever it access the channel. The distances are $d_{\text{AP,AP}} = 5$ m and $d_{\text{AP,STA}} = 2$ m for $\text{WLAN}_\text{A}$ and $\text{WLAN}_\text{C}$, and $d_{\text{AP,STA}} = 3$ m for $\text{WLAN}_\text{B}$.
		
		\item \textbf{No overlapping (Fig.~\ref{fig:x_d})}: none of the nodes causes contention to any other when transmitting. That is, every WLAN operates like in isolation. The distances in this case are $d_{\text{AP,AP}} = 10$ m and $d_{\text{AP,STA}} = 2$ m.
	\end{itemize}
	
\begin{figure}[t]
		\centering	
		\includegraphics[width=1\columnwidth]{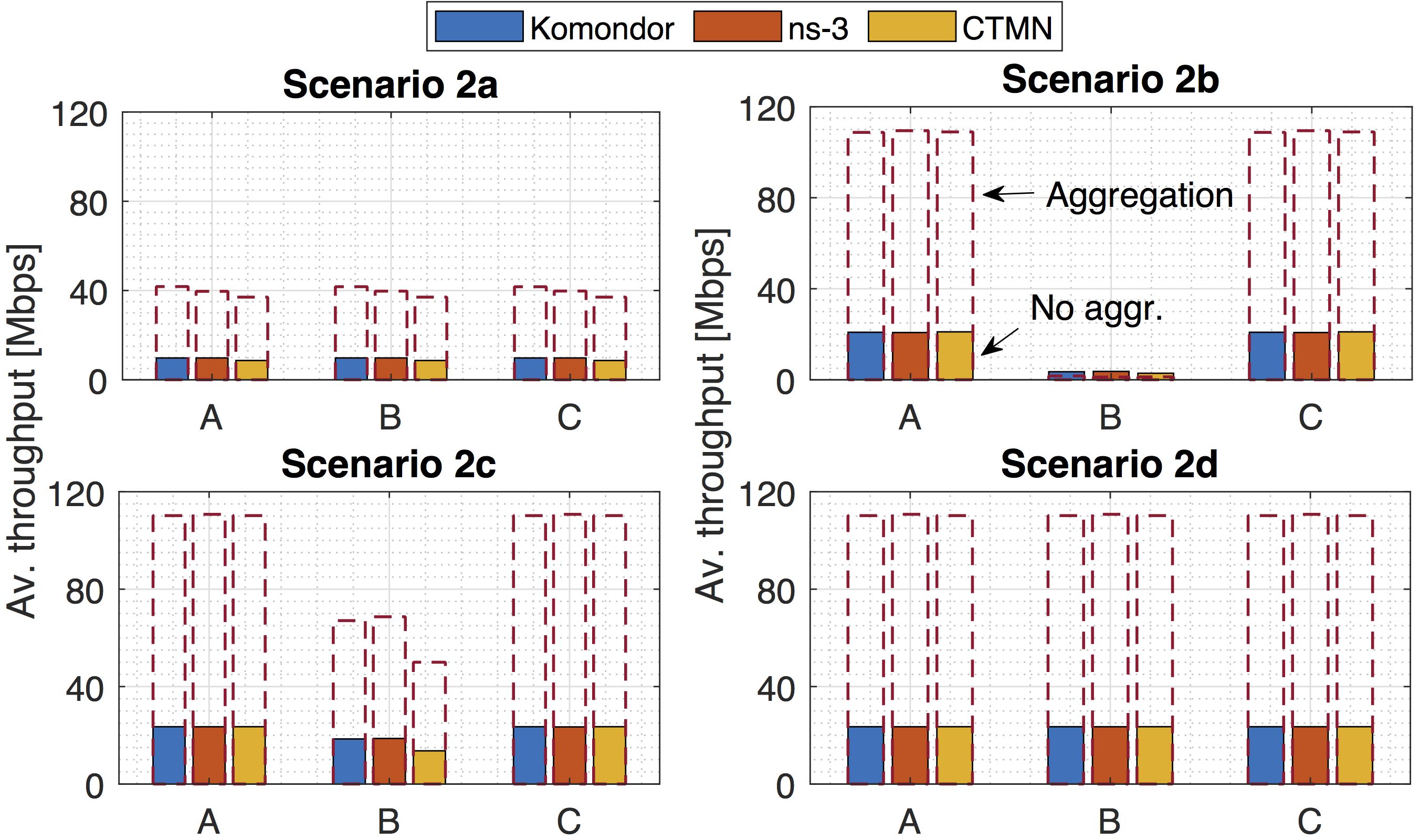}
		\caption{Average throughput experienced by each WLAN in scenarios 2a-2d. $N_{\text{agg}} = 1$ and $N_{\text{agg}}=64$ are represented through solid bars and dashed lines, respectively.}
		\label{fig:results_complex_scenarios}
	\end{figure}

	The average throughput experienced by each WLAN in each scenario is shown in Fig.~\ref{fig:results_complex_scenarios}. As previously done, we compare the performance of Komondor with ns-3 and SFCTMN. Note that results gathered by both Komondor and ns-3 are very similar in all the cases. Concerning the differences in the average throughput values estimated by both simulators and SFCTMN, we observe two phenomena with respect to backoff collisions in topologies of \textit{Scenario 2a} and \textit{2c}. First, in \textit{2a}, the throughput is slightly higher when the capture effect condition is ensured. This is due to the fact that concurrent transmissions (or backoff collisions) are permitted and captured in the simulators. Second, the most notable difference is given in \textit{2c}, which is caused by the assumption of continuous time backoffs in the CTMN. These are clear examples of the limitations of the analytical tool.
	
	\subsection{High-density and simulator performance}
	
\begin{figure}[t]
		\centering	
		\includegraphics[width=1\columnwidth]{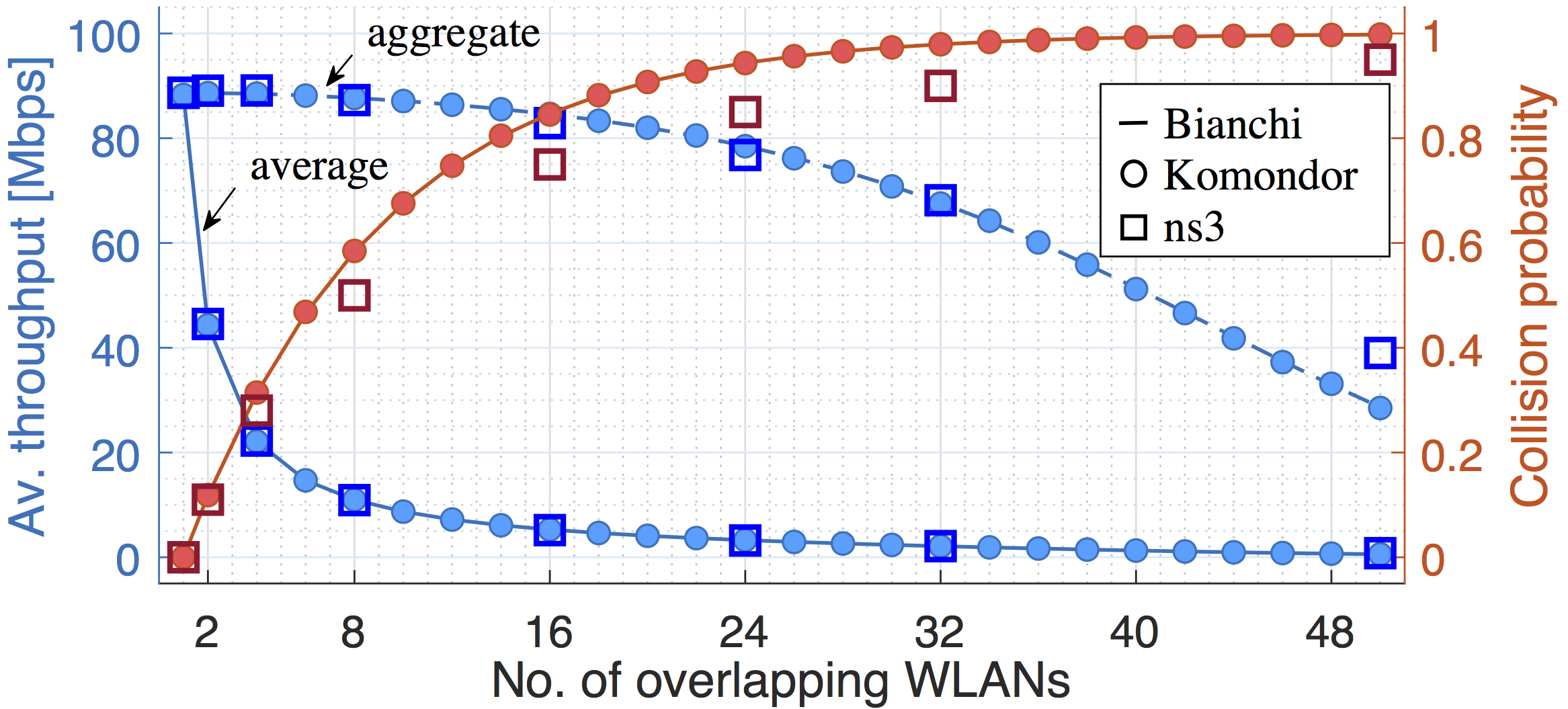}
		\caption{Throughput (average and aggregate) and collision probability vs. number of overlapping WLANs. Only some ns-3 points are plotted for the sake of visualization.}
		\label{fig:results_high_density_scenarios_throughput}
	\end{figure}

	Finally, we assess the performance of Komondor when dealing with high-density scenarios. Notice that being able to simulate scenarios with a large number of nodes is a key feature due to the ever-increasing trend towards short-range and dense deployments. In this situation, we show the results of different fully-overlapping scenarios, ranging from 1 to 50 WLANs, each consisting in of one AP and one STA. The validation is performed against the Bianchi's analytical model and ns-3. The MCS for all the WLANs is set to 256-QAM. Fig.~\ref{fig:results_high_density_scenarios_throughput} shows the results in terms of throughput (average and aggregate) and collision probability obtained for fully overlapping networks of different sizes. For comparison purposes, the simulation time used in each scenario has been set to 100 seconds, for both Komondor and ns-3. Notice that such a fully overlapping setting frames a worst-case situation regarding packet collisions. This impacts on the number of events and the simulation time as the network density increases. Nevertheless, much more positive results are expected to be achieved in more realistic non-fully overlapping dense scenarios.
	
	As shown, Komondor maintains its accuracy with respect to Bianchi's model, even when dealing with a lot of nodes. Regarding ns-3, slight differences are noticed in the collisions probability due to the error rate model, where collisions are based on the dropped RTS frames and the use of the Extended Interframe Space (EIFS). Moreover, differences in the throughput increase with the number of nodes, as previously addressed in \cite{patidar2017validation}.
	
	\label{section:density}
	
\begin{figure}[t]
		\centering	
		\includegraphics[width=1\columnwidth]{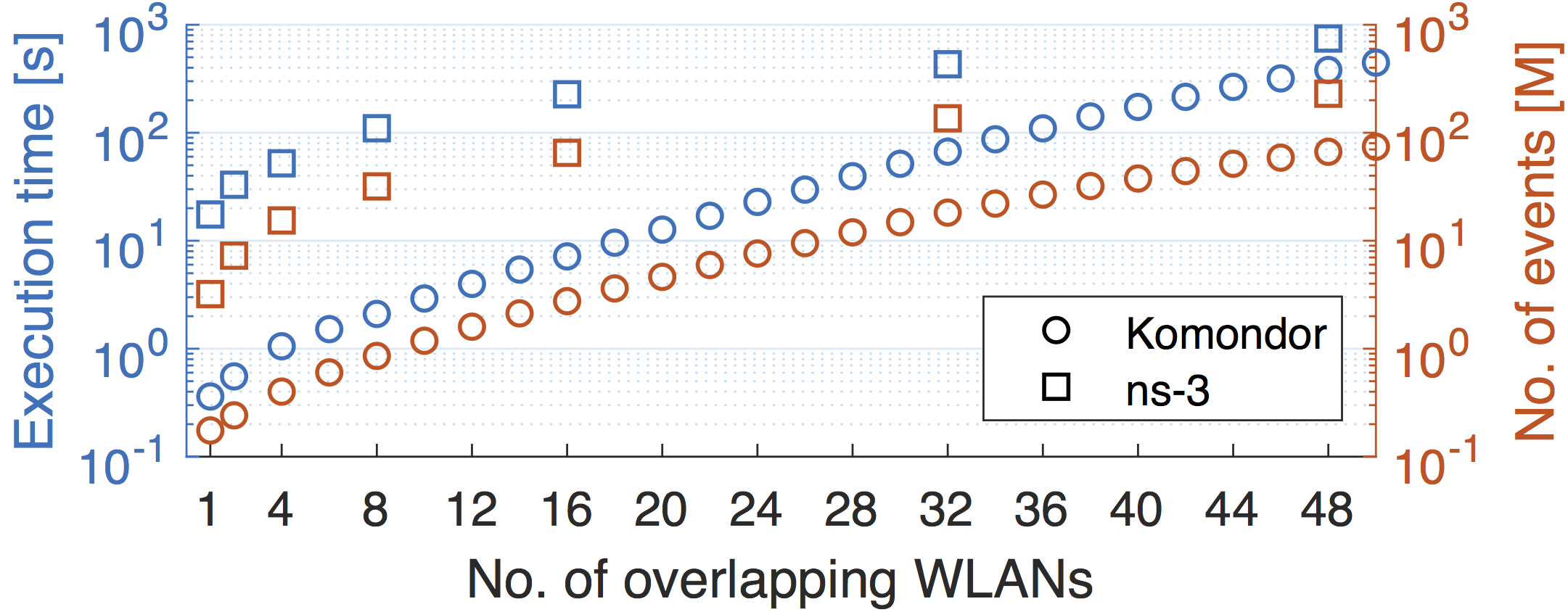}
		\caption{Execution time and number of generated events vs. number of overlapping WLANs.}
		\label{fig:results_high_density_scenarios_time}
	\end{figure}

	To conclude this section, we provide insights into the execution complexity of Komondor. Fig.~\ref{fig:results_high_density_scenarios_time} shows the execution time and the number of generated events in Komondor and ns-3 for each number of WLANs.\footnote[7]{Note that the execution time is strongly dependent of the computer used and its status at the moment of performing the simulation. In our case, we used an Intel Core i5-4300U CPU @ 1.9 GHz x 4 and 7.7 GiB memory.} As shown, the execution complexity of ns-3 is significantly higher than in Komondor. We identify the cause of this difference to be the complex PHY implementation in ns-3, which leads to a larger number of generated events.
	
	\section{Komondor and potential use cases}
	\label{section:potential}
	
	Apart from small deployments consisting of few WLANs under single-channel operation~\cite{wilhelmi2019potential}, more complex scenarios capturing DCB or high-density scenarios have been already validated and analyzed by using Komondor. In this section, we briefly discuss further potential uses such as the implementation of next-generation WLAN techniques or the inclusion of learning agents to perform efficient spectrum access and spatial reuse.
	
	\subsection{Potential usage}
	
\begin{figure}[h]
		\centering	
		\includegraphics[width=0.95\columnwidth]{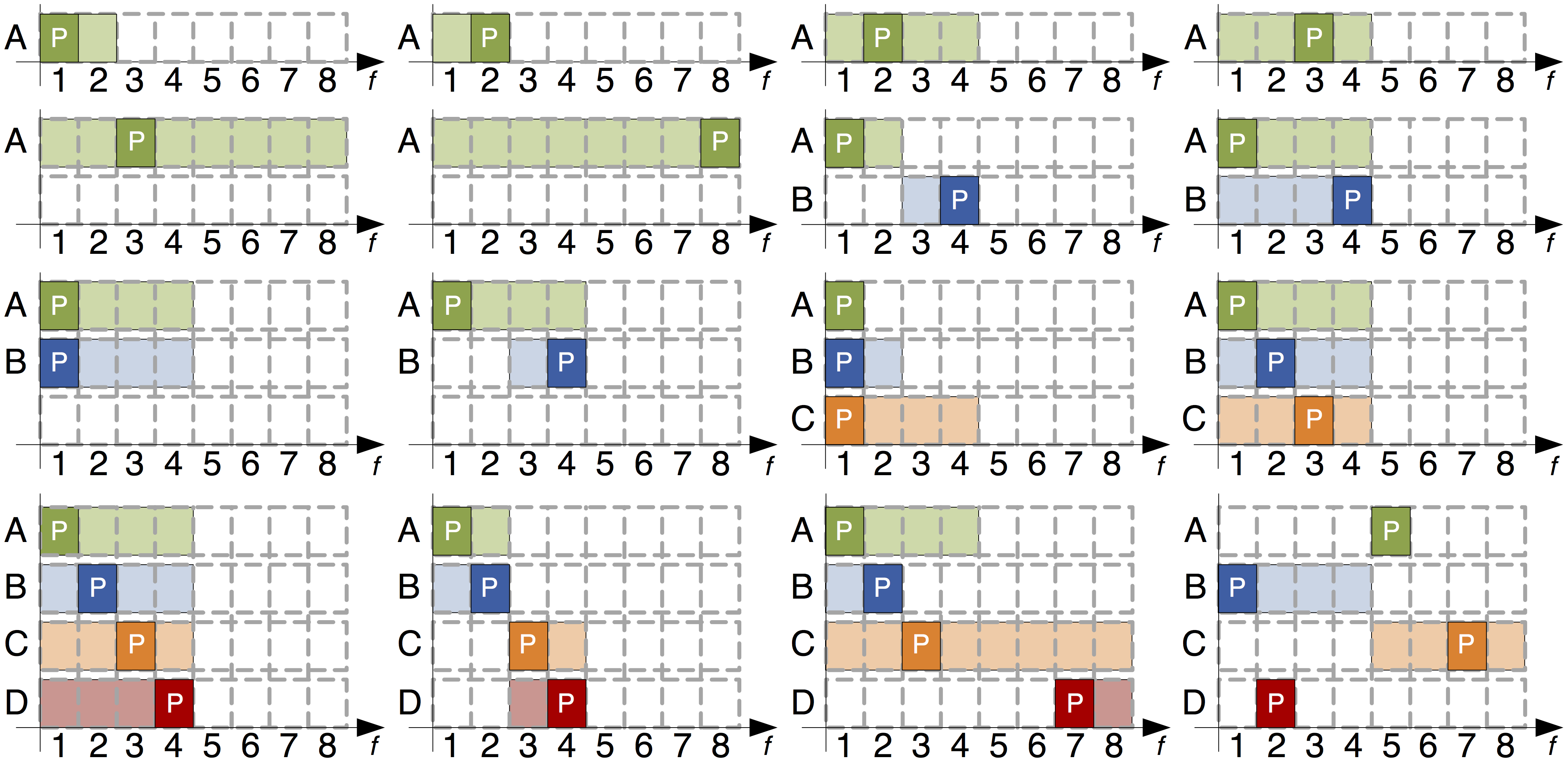}
		\caption{Scenarios with different DCB capabilities.} 
		\label{fig:channel_allocation}
	\end{figure}

	Complex wireless environments can be already extensively simulated by Komondor as a result of its reduced computational complexity in comparison to other well-known simulators such as ns-3. 	
	A prominent example of a complex scenario mixing both high-density deployments and DCB is discussed in~\cite{barrachina2019dynamic}, where authors assessed the performance of different DCB policies versus node density (see Fig.~\ref{fig:map_dense}). In~\cite{barrachina2019overlap}, a similar deployment is analyzed while considering different traffic loads.
	A set of scenarios including DCB is shown in Fig.~\ref{fig:channel_allocation}, which were validated in Komondor's validation report v0.1.\footnote[8]{Komondor's validation report v0.1: \url{https://github.com/wn-upf/Komondor/blob/master/Documentation/Other/validation\_report\_v01.pdf}.} New features like spatial reuse, MIMO, beamforming and MU communications through OFDMA and/or MU-MIMO are currently under development.
	
	\subsection{Machine learning agents}
	\label{section:machine_learning}
	
\begin{figure}[t!]
		\centering
		\begin{subfigure}[b]{0.6\columnwidth}			\includegraphics[width=0.9\textwidth]{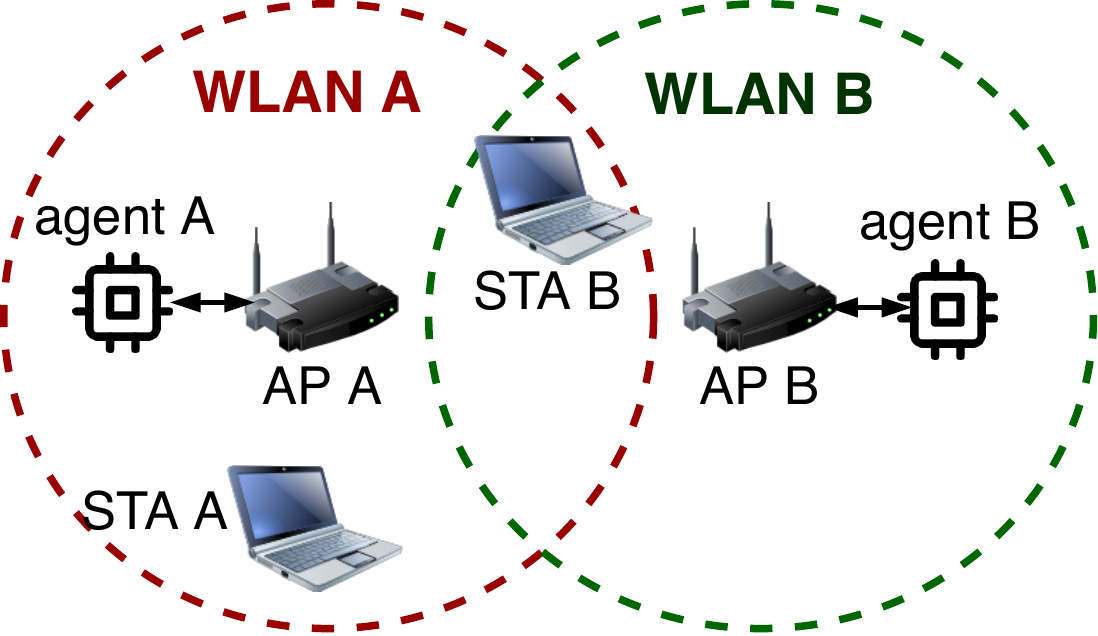}
			\caption{Agents embedded to APs}
			\label{fig:wlan_agents}
		\end{subfigure}
		\begin{subfigure}[b]{1\columnwidth}
		\includegraphics[width=\textwidth]{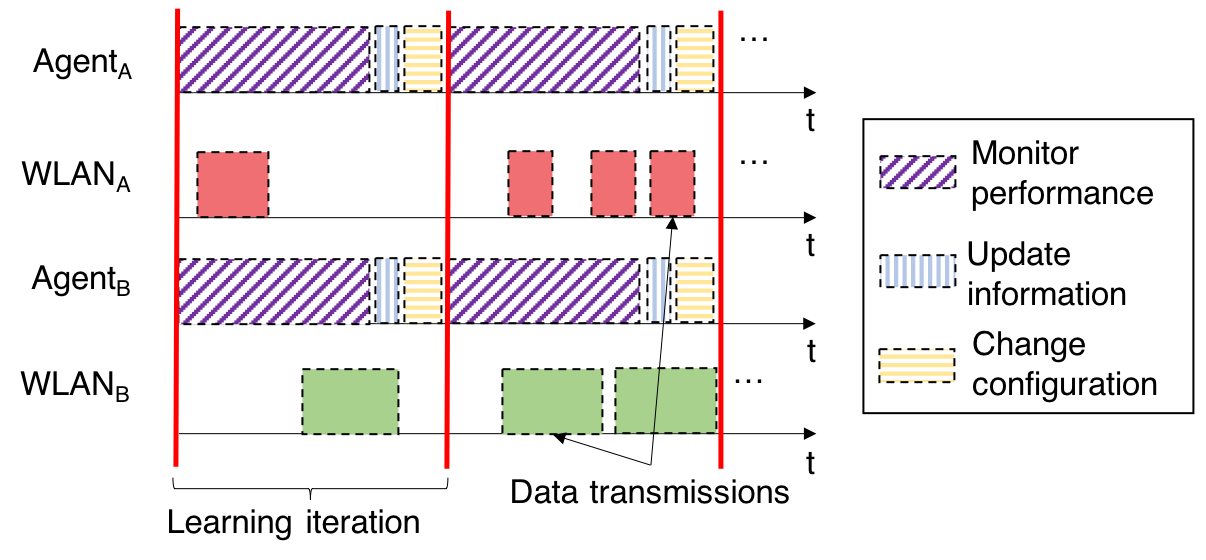}
			\caption{Learning operation followed by agents}
			\label{fig:agents_operation}
		\end{subfigure}
		\caption{ML-based operation implemented in Komondor.}
		\label{fig:agents_komondor}
	\end{figure}

	In addition to simulating advanced techniques proposed by the latest IEEE 802.11 amendments, Komondor permits including intelligent agents. In particular, agents are embedded to APs (see Fig. \ref{fig:wlan_agents}) to perform the following operations (see Fig. \ref{fig:agents_operation}): \emph{i)} monitor WLAN's performance, \emph{ii)} run an implemented learning method, and \emph{iii)} suggest new configurations to be applied by the WLAN, according to generated knowledge.
	
	The application of intelligent agents has been previously studied in \cite{wilhelmi2019collaborative, wilhelmi2019potential}, where decentralized learning is employed to both Transmit Power Control (TPC) and Carrier Sense Threshold (CST) adjustment.
	
	\section{Conclusions}
	\label{section:conclusions}
	In this work, we presented Komondor, a wireless network simulator that stems from the need of providing a reliable and low-complexity simulation tool able to capture the operation of novel WLAN mechanisms like DCB or spatial reuse. The operation of Komondor has been validated against the ns-3 simulator and analytical tools such as CTMNs and Bianchi's DCF model. In this regard, we have shown its effectiveness when dealing with high-density scenarios, thereby outperforming ns-3 with respect to the simulation time. The provided validation is fundamental for the next development stages, which contemplate the inclusion of novel techniques in WLANs that have not been fully implemented in other well-known simulators. Some future implementations contemplate OFDMA, MU-MIMO, and the spatial reuse operation, naming a few among others. Finally, we have discussed the potential of Komondor regarding complex scenarios and ML integration. In particular, a preliminary ML-based architecture is already implemented, so that intelligent agents can rule self-configuring operations at different communication levels.

	\bibliographystyle{IEEEtran}
	\bibliography{bib}
	
	\vspace{12pt}
	
\end{document}